\documentclass{emulateapj}

\usepackage{amsmath,natbib,graphicx}
\usepackage{epsf}
\usepackage{fancyvrb}
\usepackage{epstopdf}
\usepackage{epsfig}
\usepackage{color}
\DeclareGraphicsExtensions{.jpg,.pdf,.png,.eps,.ps}

\begin{document}
\VerbatimFootnotes
\title{Stabilizing Cloud Feedback Dramatically Expands 
the Habitable Zone of Tidally Locked Planets}

%\shorttitle{Clouds and the Habitable Zone}

\author{Jun Yang}
\affil{The Department of the Geophysical Sciences, The University of Chicago, 
5734 South Ellis Avenue, Chicago, IL 60637, USA}
\author{Nicolas B. Cowan}
\affil{Center for Interdisciplinary Exploration and Research in Astrophysics (CIERA) 
and Department of Physics and Astronomy, Northwestern University, 
2131 Tech Drive, Evanston, IL 60208, USA}
\and
\author{Dorian S. Abbot}
\affil{The Department of the Geophysical Sciences, The University of Chicago, 
5734 South Ellis Avenue, Chicago, IL 60637, USA}

\shortauthors{Yang, Cowan, \& Abbot}

\email{abbot@uchicago.edu}

\begin{abstract}
  The Habitable Zone (HZ) is the circumstellar region where a planet
  can sustain surface liquid water. Searching for terrestrial planets
  in the HZ of nearby stars is the stated goal of ongoing and planned
  extrasolar planet surveys. Previous estimates of the inner edge of
  the HZ were based on one dimensional radiative--convective
  models. The most
  serious limitation of these models is the inability to predict cloud behavior. Here we use global climate
  models with sophisticated cloud schemes to show that due to a
  stabilizing cloud feedback, tidally locked planets can be habitable
  at twice the stellar flux found by previous studies. This
  dramatically expands the HZ and roughly doubles the frequency of
  habitable planets orbiting red dwarf stars.  At high stellar flux,
  strong convection produces thick water clouds near the substellar
  location that greatly increase the planetary albedo and reduce
  surface temperatures. Higher insolation produces stronger substellar
  convection and therefore higher albedo, making this phenomenon a
  stabilizing climate feedback. Substellar clouds also effectively
  block outgoing radiation from the surface, reducing or even
  completely reversing the thermal emission contrast between dayside
  and nightside. The presence of substellar water clouds and the
  resulting clement surface conditions will therefore be detectable
  with the James Webb Space Telescope.
\end{abstract}

%\keywords{astrobiology --- planets and satellites: atmospheres --- stars: low-mass }


\section{Introduction}

%The traditional habitable zone (HZ) is bounded on the cold outer edge
%by the condensation of CO$_2$, and at the hot inner edge by the
%runaway greenhouse or loss of planetary water to space
%\citep{Kasting1988}.  The presence of CO$_2$ ice clouds
%\citep{ForgetandPierrehumbert1997,Wordsworthetal2011} and other
%greenhouse gases \citep{PierrehumbertandGaidos2011}
%%and potentially heat transport by oceans \citep{HuandYang2013} (I would like to statement this after the publication of the paper (JY))
%also influence the outer edge. Near the inner edge, water-vapor
%becomes a major component of the atmosphere and dominates its infrared
%opacity. Clouds would therefore have a weak greenhouse effect 
%in this regime \citep{Kasting1988}, but may significantly increase planetary 
%albedo \citep{Selsisetal2007}. Estimates of 
%%the stellar flux at 
%the inner edge of the HZ were based on 1D radiative-convective models 
%\citep{Kastingetal1993,Kopparapuetal2013}, without considering the
%effects of clouds.

Water clouds influence the climate and habitability of a planet by
either scattering incoming stellar radiation back to space (cooling)
or by absorbing and reradiating thermal emission from the surface
(warming). Near the cold outer edge of the HZ \citep{Kastingetal1993},
CO$_2$ ice clouds can scatter outgoing thermal radiation back
to the surface, also causing warming
\citep{ForgetandPierrehumbert1997, Wordsworthetal2011}. Near the hot
inner edge of the HZ, water vapor becomes a major component of the
atmosphere and dominates its infrared opacity. Water clouds would
therefore have a weak greenhouse effect in this regime
\citep{Kasting1988}, but may significantly increase planetary albedo
\citep{Selsisetal2007,Kalteneggeretal2011}.

Previous investigations of the inner edge of the HZ have typically
been based on 1D radiative-convective models that neglect cloud effects
\citep{Kastingetal1993,Kopparapuetal2013}.  Recently, one dimensional
(1D) cloud models were developed to investigate the effects of clouds
on exoplanet climate \citep{Kitzmannetal2010,Zsometal2012}.  These
models incorporate a treatment of cloud microphysics, but cannot
predict cloud coverage, location, or altitude, all of which are
essential for determining cloud radiative effects. Such cloud features
can only be predicted by making 3D dynamical calculations of
atmospheric circulation. A few studies with 3D general circulation
models (GCMs) have simulated clouds of tidally locked planets at low
stellar fluxes \citep{Joshi2003,Edsonetal2011,Edsonetal2012}, but
the impact of 3D cloud behavior on the inner edge of the HZ has not
been considered. Here we study this issue using GCMs that explicitly
calculate atmospheric dynamics, radiative transfer, the hydrological
cycle, and clouds. %formation.

%Clouds influence the climate and habitability of a planet by
%either scattering incoming stellar radiation back to space (cooling)
%or by absorbing and reradiating thermal emission from the surface
%(warming). 

%The position of the outer edge of the HZ is poorly defined and can be
%affected by carbon dioxide clouds
%\citep{ForgetandPierrehumbert1997,Wordsworthetal2011}, the presence of
%other greenhouse gases \citep{PierrehumbertandGaidos2011}, and potentially
%heat transport by oceans \citep{Huetal2013}. The inner edge of the HZ
%is based on entirely different physical processes than the outer edge,
%and has traditionally been calculated using 1D models. Estimates of
%the stellar flux at the inner edge of the HZ have been consistent to
%within $\approx$10\% over the past twenty years
%\citep{Kastingetal1993,Kopparapuetal2013}. These calculations, however,
%neglect the effects of water clouds, which can influence the climate
%of a planet by either scattering incoming stellar radiation back to
%space (cooling) or by absorbing and reradiating thermal emission from
%the surface (warming).

%This makes it sound like the inner edge is just a wishy-washy as the outer edge. NBC

We focus on planets around M dwarf stars, which have small masses,
low photospheric temperatures, and constitute $\approx$75\% of
main sequence stars. 
%\citep{Pierryman2011}.
Current data suggest there is $\approx$1 Earth-size planet in the HZ
per M-star \citep{DressingandCharbonneau2013,MortonandSwift2013}, and
these planets are relatively easy to detect by radial velocity and
transit surveys \citep{CharbonneauandDeming2007}. Because tidal forces
are strong at short distances, habitable planets in circular orbits around low-mass
stars are expected to be tidally locked, with one
hemisphere experiencing perpetual day and the other perpetual night.

%in circular orbits 

%As we will describe below, a combination of higher cloud albedo
%and reduced greenhouse effect leads to dramatically lower surface
%temperatures in tidally locked than non-tidally locked GCM simulations. 
%Consequently, the tidally locked simulations produce stable, habitable 
%climates at approximately double the stellar flux that defines the 
%inner edge of the HZ in 1D model calculations. Since these GCMs become 
%numerically unstable at temperatures much lower than the 
%runaway greenhouse limit, the highest stellar flux at which the 
%GCMs converge is a conservative estimate of the inner edge. 
%Our results therefore require a significant expansion of the HZ
% to include much higher stellar fluxes
%for tidally locked planets.
% This is redundant with Abstract & Conclusions. NBC

%%%%%%%%%%%%%%%%%%%%%%%%%%%%%%%%%%%
%%%%%%%%%%%%%%%%%%%%%%%%%%%%%%%%%%%

%\clearpage
\begin{figure*}[]
\vspace{-45mm}
\begin{center}
\includegraphics[angle=0, width=40pc]{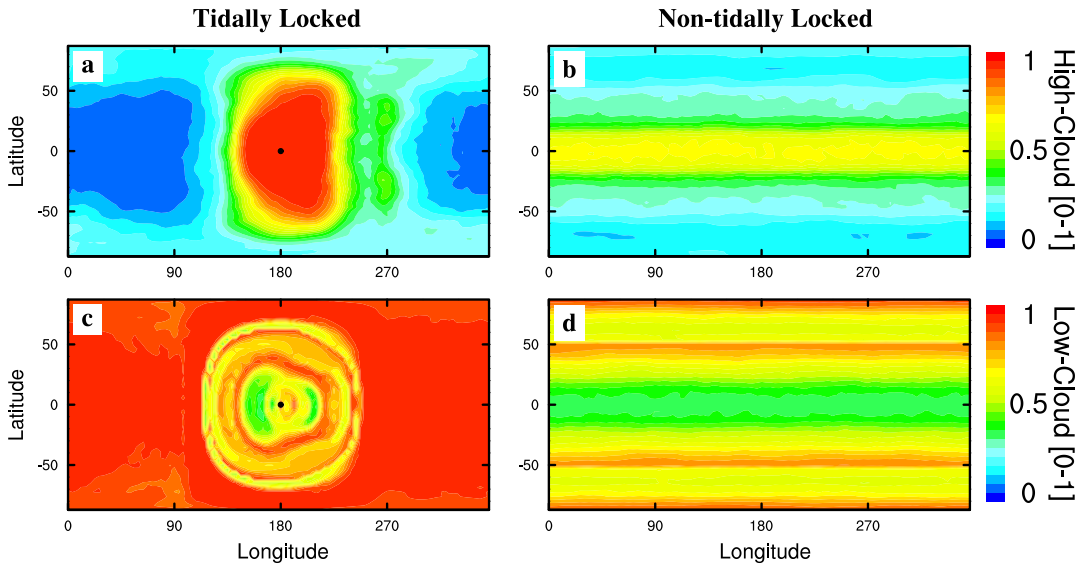}
\end{center}
\vspace{-50mm}
\caption{Cloud behavior for tidally (left) and non-tidally (right)
  locked planets. High-level cloud fraction (top) and low-level cloud fraction
  (bottom) are displayed in each case. The non-tidally locked case is
  in a 6:1 spin--orbit resonance. The stellar flux is
  1400\,W\,m$^{-2}$. The black dot in (a\,\&\,c) is substellar
  point.}
\label{fig1}
\end{figure*}

\section{Global Climate Models}
We perform simulations using 3D GCMs with sophisticated cloud
schemes. The climate models we use include atmospheric GCMs: the
Community Atmosphere Model version 3.1
\citep[CAM3;][]{Collinsetal2004}, version 4.0
\citep[CAM4;][]{Nealeetal2010a}, and version 5.0
\citep[CAM5;][]{Nealeetal2010b} coupled with a mixed layer (immobile)
ocean with a uniform depth of 50 meters, and a fully coupled
atmosphere-ocean GCM: the Community Climate System Model version 3.0
\citep[CCSM3;][]{Collinsetal2006} with a uniform ocean depth of 4000
m.  All the GCMs simulate marine stratus, layered clouds, shallow and
deep convective clouds in both liquid and ice phases.
%, and are able to consistently predict cloud water path, cloud altitude and cloud coverage. 
CAM4 has a new deep-convection scheme relative to CAM3, and 
CAM5 has a completely new cloud parameterization that is different 
from CAM3 and CAM4. CAM5 attempts to simulate full aerosol--cloud interactions.

CAM3 solves the equations for atmospheric motion on a rotating sphere
and the equations for radiative transfer including the effects of
water vapor, greenhouse gases and clouds. The radiative scheme is
accurate for atmospheres with CO$_2$ concentration up to $\approx$0.1~bar
\citep{Abbotetal2012b} and water vapor column content less than
1200\,kg\,m$^{-2}$ \citep{Pierrehumbert2010}; in all our simulations
CO$_2$ and water vapor are below these limits. Cloud absorption and shortwave 
scattering are represented in terms of cloud water content, cloud
fraction and droplet effective radius.
% for four spectral intervals between 0.25 and 5.0\,$\micron$. 
Cloud infrared scattering is not included as it is
  negligible \citep{Fuetal1997} due to high infrared absorption by
  water clouds \citep{Pierrehumbert2010}. Cloud fraction is
parameterized based on relative humidity, atmospheric stability and
convective mass fluxes. Cloud water content is determined
prognostically by a cloud microphysical scheme.

%We run the
%model at a horizontal resolution of 3.75$^{\circ}$$\times$3.75$^{\circ}$
%and with 26 vertical levels from the surface to $\approx$30 km. 

We have modified the models to be able to simulate the climates of 
extrasolar planets with different stellar spectra, orbits, and 
atmospheres. The stellar spectrum we use is for an M-star (or K-star) with 
an effective temperature of 3400~(or~4500)~K, assuming a blackbody 
spectral distribution. The stellar flux is set to a sequence of values 
from~1000~to~2600\,W\,m$^{-2}$, which corresponds to moving 
the planet closer to the central star. The geothermal heat flux is set to zero.
By default, the radius, gravity and orbital period of the planet are 
set to 2~$R_\oplus$ ($R_\oplus$ is Earth radius), 1.4~$g_\oplus$ 
($g_\oplus$ is Earth gravity) and 60~Earth days (E-days), respectively. 
Both the obliquity and eccentricity are set to zero. For tidally locked 
simulations, the rotation period is equal to the orbital period (1:1). 
For non-tidally locked simulations, the spin-orbit resonance is set to 
2:1 (or 6:1), meaning that the rotation frequency is 2~(or~6) times 
faster than the orbital frequency.

The default atmosphere is 1.0~bar of N$_2$
and H$_2$O.  Assuming the silicate weathering hypothesis
\citep{Walkeretal1981} is correct, the CO$_2$ level should be low near
the inner edge of the HZ, although this depends on the presence and
location of continents \citep{Edsonetal2012,Abbotetal2012a}.  We
therefore set CO$_2$ to zero, as well as CH$_4$, N$_2$O, and O$_3$. 
Sensitivity tests with different CO$_2$
levels show that this has little effect on our results
(Section~\ref{Sensitivity}). The experimental design for CAM4 and CAM5
is the same as CAM3 except that aerosols are set to zero in CAM3 and
CAM4 but to 20$^{th}$\,century Earth conditions in CAM5 because its
cloud scheme requires aerosols.

%We ran CAM4 and CAM5 at a horizontal resolution of 
%1.9$^{\circ}$$\times$2.5$^{\circ}$ and a vertical resolution of 26
%and 30 levels, respectively. 
%All the experiments were initialized 
%from a uniform surface temperature of~280\,K. The model default 
%time step is 1800~s; for high stellar flux, it is reduced to~100~or~50 s 
%to avoid numerical instability. Each case was integrated for 65~E-years 
%and reaches steady-state within 30~E-years. 
%%The last 5 years are employed for the analyses. 
%
% 
 %The stellar flux is set to 1600\,W\,m$^{-2}$, which 
 %is the flux at which the stabilizing cloud feedback is active in CAM3 simulations.
 %and there is a substantial difference between the tidally locked and non-tidally locked CAM3 simulations. 
 %For illustrative purposes, we assume the ocean has a uniform 
 %depth of 4000~m (the mean value on modern Earth). 
 % In order to know the effect of ocean dynamics on the mechanism we describe in 
 % hypothetical fairly wet planets, 
We consider three idealized continental configurations in our
ocean-atmosphere CCSM3 simulations: (i) aqua-planet: a globally ocean
covered planet; (ii) one ridge: a thin barrier that completely
obstructs ocean flow running from pole to pole on the eastern
terminator; (iii) two ridges: two thin barriers on the western and
eastern terminators, respectively. The planetary parameters for the
CCSM3 simulations are a 37~E-days orbit period, a radius of
1.5~$R_\oplus$, and gravity of 1.38~$g_\oplus$ around an M-star.

\section{The Stabilizing Cloud Feedback}

\begin{figure*}[]
\vspace{-12mm}
\includegraphics[angle=0, width=40pc]{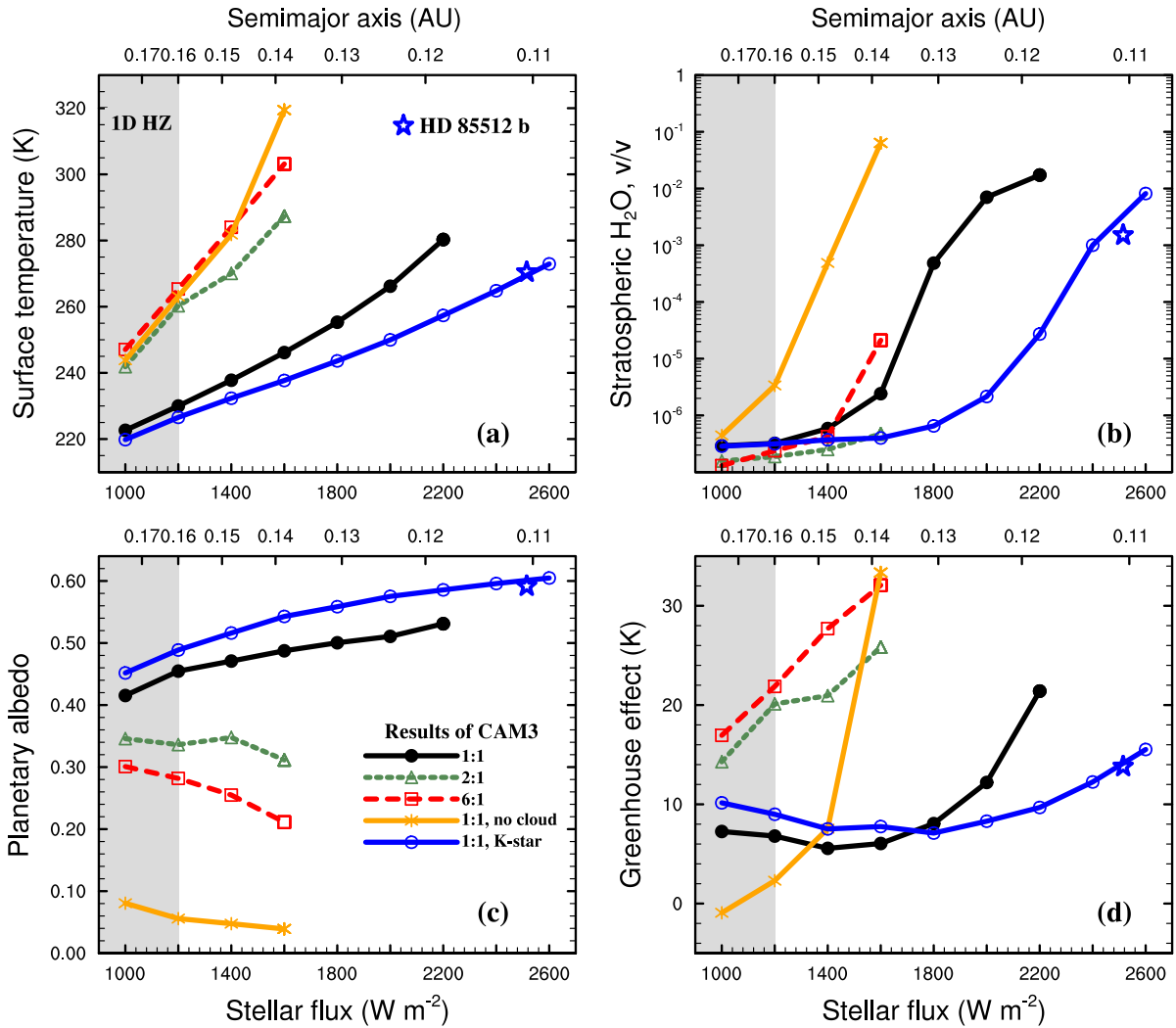}
\vspace{-15mm}
\caption{Climates of tidally locked and non-tidally locked terrestrial planets. 
(a) global-mean surface temperature (K), 
\textcolor{black}{(b) stratospheric H$_2$O volume mixing ratio at the substellar point}, 
(c) planetary albedo and (d) global-mean greenhouse effect (K).
%, as a function of the stellar flux. 
The upper horizontal axis is the corresponding semimajor axis between an M-star 
(with a 2.3\% $L_\odot$) and the planet. %in AU. 
%1\,AU is the Earth--Sun distance. 
1:1 denotes a tidally locked state, %(one rotation per orbit) 
and 2:1 and 6:1 denote 2 or 6 rotations per orbit, respectively. 
For ``no cloud'' cases, all clouds are set to zero. 
The stellar spectrum is for an M-star or an K-star. 
Results for HD\,85512\,b ($S_0$\,=\,2515~W\,m$^{-2}$, 1.5 $R_\oplus$, 1.6 $g_\oplus$ 
and a 58 Earth-day orbit) are represented by a pentagram. 
%Its planetary parameters are $S_0$\,=\,2515~W\,m$^{-2}$ around an K-star, 
%a 58 Earth-day orbit, a radius of 1.5 $R_\oplus$, and gravity of 1.6 $g_\oplus$. 
The gray area denotes the HZ around an M-star with an
inner edge of $S_0$$\approx$1200~W\,m$^{-2}$ and an outer edge of
$S_0$$\approx$270~W\,m$^{-2}$ (not shown), obtained in a 1D model
without clouds \citep{Kopparapuetal2013}.}
\label{fig2}
\end{figure*}

For tidally locked planets, most of the dayside is covered by clouds
(Fig.\,\ref{fig1}) with a high cloud water content (several times
Earth's), arising from near-surface convergence and resulting
convection at the substellar point. This circulation produces
high-level and low-level clouds covering $\approx$60\% and 80\% of the
dayside, respectively. Since thick clouds occur right where the
insolation is highest, they significantly increase the planetary
albedo. Furthermore, the planetary albedo increases with the stellar
flux ($S_0$) for planets around both M- and K-stars
(Fig.\,\ref{fig2}), leading to a stabilizing cloud feedback on
climate. This is due to increased convection at the substellar point
as surface temperatures there increase.  For an insolation of
2200~W\,m$^{-2}$ and an M-star spectrum, the planetary albedo reaches
0.53,
%the global-mean surface temperature (T$_S$) is 281~K, 
and the maximum surface temperature on the planet is only 305~K. 
The greenhouse effect from clouds is similar to that on
Earth, with a global-mean value of 20-30~W\,m$^{-2}$. For a hotter
K-star, the stellar spectrum is such that the planetary albedo is
higher at all stellar fluxes and reaches 0.60 for
$S_0$\,=\,2600~W\,m$^{-2}$. Assuming HD\,85512\,b \citep{Pepeetal2011}
is tidally locked and Earth-like, our calculations suggest it has an
albedo of 0.59 and a mean surface temperature of 271~K. 
The recently discovered super-Earth GJ\,163\,c \citep{Bonfilsetal2013}, 
provided it is tidally locked, should also have a high albedo and therefore be habitable.

%, Kalteneggeretal2011

The cloud behavior for non-tidally locked planets contrasts markedly
with that of tidally locked planets. Non-tidally locked planets have
an albedo similar to Earth's ($\approx$0.3).  This is because only
part of the tropics and the mid-latitudes are covered by clouds
(Fig.\,\ref{fig1}) and the cloud water content is relatively small. In
contrast to the tidally locked case, the planetary albedo decreases as
$S_0$ is increased (Fig.\,\ref{fig2}c), representing a destabilizing
feedback. This is primarily due to decreases in cloud coverage
associated with reduced latitudinal temperature gradient and resulting
weakened Hadley cells.

In order to clearly demonstrate the influence of clouds, we switch off
the cloud module in the model and repeat the tidally locked
simulations.  At an insolation of 1600~W\,m$^{-2}$, the planetary
albedo drops from~0.53~to~0.04, and the global-mean surface
temperature rises from~246~to~319~K (Fig.\,\ref{fig2}a). Clouds
therefore account for 73~K of cooling, which is critical for extending
the HZ to higher stellar flux.

%\clearpage
\begin{figure*}[]
\vspace{0mm}
\includegraphics[angle=0, width=40pc]{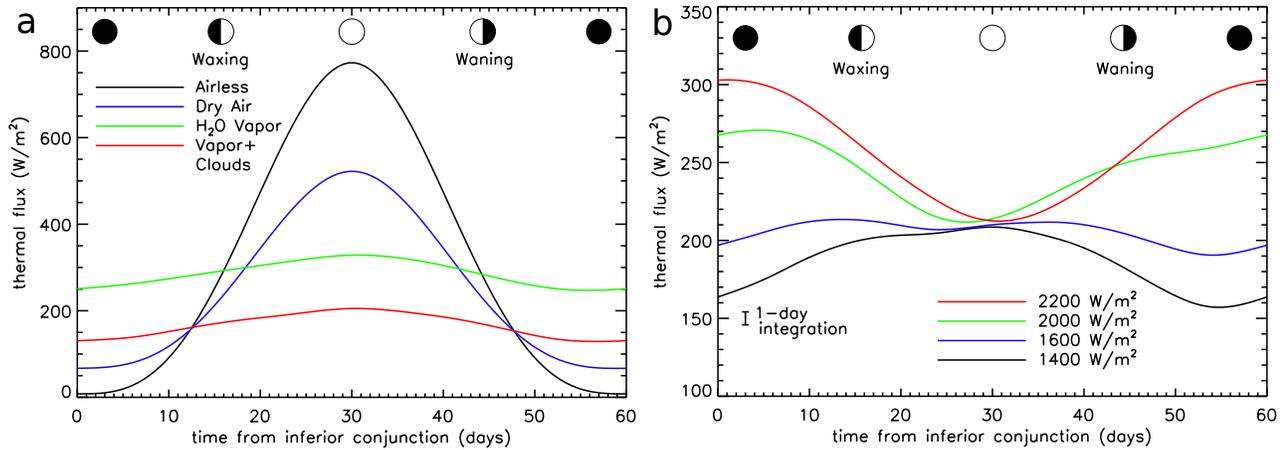}
\vspace{0mm}
\caption{Thermal phase curves of tidally locked planets. 
(a) phase curves for different atmospheres with stellar
flux fixed at 1200\,W\,m$^{-2}$:~airless, dry-air, water vapor, and
water vapor plus clouds, (\textbf{b}) phase curves for a full atmosphere
including water vapor and clouds for different stellar fluxes:\,1400,
1600, 2000 and 2200\,W\,m$^{-2}$. The error bar in (b) is the expected
precision of the James Webb Space Telescope for observations of a nearby super-Earth. 
The surface albedo for the airless and dry-air cases is 0.2. 
The orbital period is 60 Earth-days.}
\label{fig3}
\end{figure*}

Tidally locked planets not only have a higher cloud albedo, but also a
smaller greenhouse effect ($G=T_{em}^{srf}-T_{ef}$, where
$T_{em}^{srf}=\left(\frac{F^{srf}_\uparrow}{\sigma}\right)^\frac{1}{4}$
and $T_{ef}=\left(\frac{F^{top}_\uparrow}{\sigma}\right)^\frac{1}{4}$,
where $F^{srf}_\uparrow$ and $F^{top}_\uparrow$ are the global mean
upward infrared radiative fluxes at the surface and top of atmosphere,
respectively).  For reference, $G \approx$33~K on modern Earth.  The
greenhouse effect for tidally locked planets is much smaller than
those for non-tidally locked planets (Fig.\,\ref{fig2}d).  This
results from a low-level temperature inversion on the nightside
%and in the extra-tropics of the dayside (For high stellar flux, the inversion on the dayside may be weak.) (JY)
of tidally-locked planets \citep[see
also][]{Joshietal1997,Leconteetal2013}. The inversion is due to
efficient radiative cooling by the surface on the nightside and strong
atmospheric energy transport from the dayside to the nightside
\citep{MerlisandSchneider2010}. The outgoing
infrared radiation to space %emitted by the atmosphere
is therefore similar to the near-surface upward infrared radiation, resulting in a
small $G$.

%\textcolor{red}{Fig.~2b shows that when the stellar flux is $\geq$2200~W\,m$^{-2}$, 
%the stratospheric water-vapor mixing ratio is close to or slightly higher 
%than the limit for the moist greenhouse effect, $\approx$3.0$\times$10$^{-3}$, 
%as suggested by Kasting et al. (1993). This limit, however, is poorly constrained, and 
%depends on ocean-water volume, atmospheric composition, the edge and gravity of the planet. 
%Future research with a model including the processes of H$_2$O photolysis and of 
%hydrogen escape would be required to further study 
%the moist greenhouse effect on tidally locked planets 
%with the stellar flux we discussed here.} 

Our results demonstrate that at high stellar flux, Earth-like tidally
locked planets have a high planetary albedo, a low greenhouse effect,
and therefore low enough surface temperatures to be habitable.  
The sub-stellar and global-mean surface temperatures in our simulations 
remain well below the 340~K moist greenhouse limit of \cite{Kopparapuetal2013}, 
but the dayside stratospheric water-vapor mixing ratio in our hottest simulations (Fig.~2b)
is comparable to the $\approx$3.0$\times$10$^{-3}$ 
moist greenhouse limit suggested by \cite{Kastingetal1993}. 
Future modeling including H$_2$O photolysis and 
hydrogen escape would be need to determine 
the moist greenhouse threshold for tidally locked planets.

%Since
%GCMs become numerically unstable \textcolor{red}{as the atmosphere is close to the 
%moist greenhouse limit (Fig.~2b)}, the highest stellar flux at which our
%simulations converge is a conservative estimate of the inner edge of
%the HZ.  

%Sensitivity tests with CAM3 and further simulations with
%CAM4, CAM5 and CCSM3 show that our results are robust (Section 5).

%%%%%%%%%%%%%%%%%%%%%%%%%%%%%%%%%%%
%%%%%%%%%%%%%%%%%%%%%%%%%%%%%%%%%%%

\section{Thermal Phase Curves}\label{phases}
The key to observationally confirming the presence of substellar clouds
is that they also affect the planet's outgoing longwave radiation
(OLR). Following \citet{Cowanetal2012b}, we use the OLR maps from the
GCMs to compute time-resolved, disk-integrated broadband thermal phase
curves measured by a distant observer (Fig.~\ref{fig3}).  Interannual
variability in OLR is less 1\% for our simulations; we
therefore use the long-term mean maps. The sub-observer latitude is
set to zero. Although this corresponds to an edge-on orbit for a
zero-obliquity planet, the transits and eclipses have been omitted for clarity.

%To introduce this idea, we consider the 
%effects of increasingly complex atmospheres on OLR. A tidally locked 
%airless planet has an OLR maximum at the substellar point.
%%\citep{Maurinetal2012}. 
%If a tidally locked planet has a dry atmosphere, equatorial 
%superrotation causes an eastward displacement of the OLR maximum 
%from the substellar point by 2$^\circ$-4$^\circ$ longitude , 
%which is  When we include 
%water-vapor, but not clouds, the OLR maximum shows a \emph{westward} 
%displacement from the substellar point by about 50$^\circ$ (Fig.\,3e). 
%This is due to the eastward displacement of the water-vapor concentration 
%maximum (Fig.\,3c) induced by the equatorial superrotation, and the decrease 
%in OLR on the eastern side of the substellar point caused by strong
%water-vapor absorption. If clouds are also included, OLR decreases
%everywhere due to the lower surface temperatures (Figs.\,3a-b), and an
%OLR \emph{minimum} appears at the substellar point (Fig.\,3f). 
%The minimum occurs because high-altitude clouds (Fig.\,1a) absorb 
%infrared radiation from the underlying atmosphere and re-emit to space 
%at their temperature, which is very low.

An airless planet exhibits large-amplitude thermal phase variations
with a peak at superior conjunction \citep[Fig.~\ref{fig3}a; see also][]{Maurinetal2012}. For the
dry-air case, the peak thermal emission occurs slightly before
superior conjunction because of equatorial superrotation \citep[see
also][]{Selsisetal2011}. This is qualitatively similar to the eastward
displacement of 5$^\circ$-60$^\circ$ typical for hot Jupiters
\citep{Knutsonetal2007,Crossfieldetal2010, Cowanetal2012a,
  Maxtedetal2013}. Equatorial superrotation is a generic feature of
tidally locked planets arising from
strong day--night forcing \citep{ShowmanandPolvani2011}.

\begin{figure*}[]
\vspace{-15mm}
\includegraphics[angle=0, width=40pc]{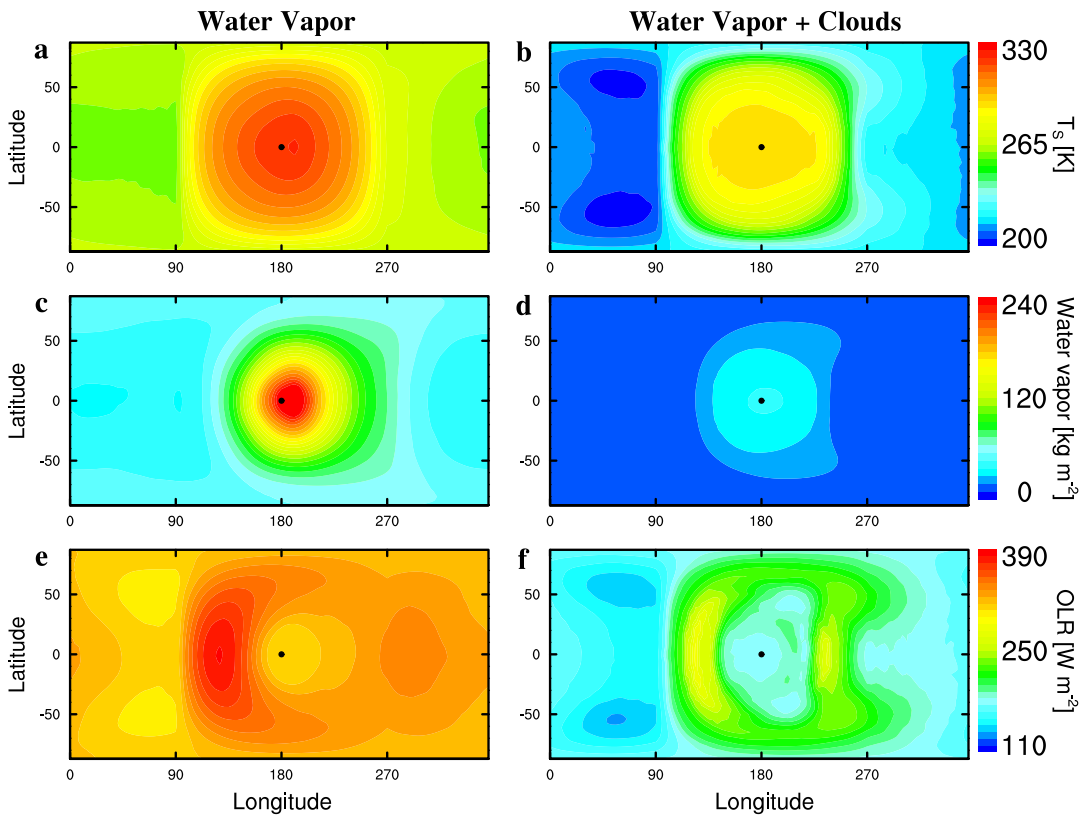}
\vspace{-48mm}
\caption{Water vapor and cloud behavior on a tidally locked
  terrestrial planet. Left panels: clouds are set to zero; right
  panels: clouds are interactively calculated. (a) and (b) surface
  temperature (T$_S$, K), (c) and (d) vertically integrated
  water vapor amount (kg\,m$^{-2}$), (e) and (f) outgoing longwave
  radiation at the top of the atmosphere (OLR, W\,m$^{-2}$). 
  The stellar flux is
  1400~W\,m$^{-2}$ for all plots.  The black dot is substellar
  point.}
  \label{fig4}
\end{figure*}

Water vapor and clouds, however, dramatically modify the phase
curves. Water vapor is advected east of the substellar point
(Fig.~\ref{fig4}), where it absorbs outgoing thermal radiation,
which leads to a lightcurve maximum \emph{after} superior conjunction
(Fig.~\ref{fig3}b). Substellar clouds also absorb thermal emission,
producing a local minimum in OLR (Fig.~\ref{fig4}f) that weakens or
completely reverses the day-night thermal flux contrast (Fig.~\ref{fig3}b). 
Note that as the stellar flux is varied, the thermal
flux near superior conjunction stays roughly constant and equal to the
high-level cloud emission flux, which corresponds to the temperature
at the tropopause. If convective
clouds occur on hot Jupiters, the interpretation of their thermal
phase variations \citep[e.g.,][]{Knutsonetal2007} will have to be
revisited.

To determine whether these phase curve features would be detectable by
the James Webb Space Telescope (JWST), we estimate the expected
signal-to-noise ratio. The contrast ratio over some band $[\lambda_1,
\lambda_2]$ is given by
\begin{equation}\label{signal}
\frac{F_p}{F_*} = \left(\frac{R_p}{R_*} \right)^2 \frac{\int_{\lambda_1}
^{\lambda_2} B(T_{ef},\lambda) d\lambda}{\int_{\lambda_1}^{\lambda_2}
 B(T_{*},\lambda) d\lambda},
\end{equation}
assuming that both the star and the planet emit as blackbodies. 
Adopting values of $R_p$\,=\,$2R_\oplus$, $R_*$\,=\,$0.2R_\odot$, 
$T_{ef}$\,=\,$240$~K, and $T_*$\,=\,$3000$~K, %yields a planet--star 
%flux ratio of $6.8$$\times$$10^{-4}$ on the Rayleigh-Jeans tail 
%of the spectral energy distributions. At the peak 
%of the planetary spectral energy distribution, $\approx$12~$\micron$, 
%the contrast is only $2.9$$\times$$10^{-5}$, while 
the band-integrated (10--28~$\micron$) contrast for the Mid-Infrared Instrument on JWST is $4.9$$\times$$10^{-5}$.

%stellar and planetary 

To estimate the photometric precision one can expect for JWST, 
we begin with the $4$$\times$$10^{-5}$ precision obtained at 3.6~$\micron$ 
for HD\,189733 with the Spitzer Space Telescope \citep{Knutsonetal2012}. 
We then scale this by stellar flux and distance, integration time, 
and telescope size, assuming Poisson noise, $\sigma/F_* \propto1/\sqrt{N}$. 
The number of photons is
\begin{equation}\label{noise}
N = \pi D^2 \tau \left(\frac{R_*}{d}\right)^2 \int_{\lambda_1}
^{\lambda_2} \frac{B(T_{*},\lambda)}{E(\lambda)} d\lambda
\end{equation}
where $E(\lambda)$\,=\,$hc/\lambda$ is the energy per photon. 
We adopt the same values as above, in addition to $D$\,=\,$6.5$~m, $\tau$\,=\,$1$\,day, and $d$\,=\,$5$ or 20~pc \citep[conservative distances to the nearest non-transiting and transiting M-Dwarf HZ terrestrial planets, respectively;][]{DressingandCharbonneau2013}. The  
expected photometric precision is $2.5\times10^{-7}$ (non-transiting)
and $1.0\times10^{-6}$ (transiting). Note that the unknown radius and orbital inclination of non-transiting planets 
\citep[][]{Cowanetal2007,Crossfieldetal2010} is not an obstacle 
for the current application because it is the lightcurve morphology, 
rather than amplitude, that betrays the presence of substellar clouds.

%Non-transiting planets exhibit thermal phase variations 
%that are attenuated by the sine of the orbital inclination, but 
%we expect non-transiting targets to be closer than their transiting counterparts. 

Dividing the contrast ratio by the precision estimates, we find that a 1-day 
integration with JWST could produce a 49$\sigma$ detection
 of broadband planetary emission for a nearby transiting M-dwarf
 HZ planet (error bar in Fig.\,\ref{fig3}b).

%The James Webb Space Telescope should be able to measure the 
%phase curves of terrestrial planets orbiting nearby red dwarfs at 
%sufficient precision (Appendix A) to validate or refute the substellar cloud
%effect we describe here.

%%%%%%%%%%%%%%%%%%%%%%%%%%%%%%%%%%%
%%%%%%%%%%%%%%%%%%%%%%%%%%%%%%%%%%%

%\clearpage
%\doublespacing

\section{Sensitivity Tests}\label{Sensitivity}
We have performed sensitivity experiments over a wide range of
parameters, including cloud particle sizes, cloud
fraction parameters, surface pressure, CO$_2$ concentration, 
planetary radius, rotation rate, surface gravity, oceanic mixed
layer depth, convective scheme, model resolution, land-sea
distribution and ocean heat transport. Results are shown in Table~1. 
Both the planetary albedo and the spatial pattern of OLR are 
insensitive to changes in almost all of these parameters.
%, which indicates that the basic mechanism of
%the negative cloud feedback that we describe is robust.  
%The spatial pattern of OLR
%, which determines the thermal phase curve and
%detectability of the negative cloud feedback, 
%is also insensitive to changes in almost all of these parameters. 
In particular, most of our tests exhibit a high albedo and significant
OLR minimum near the substellar point, especially at high stellar
flux. Furthermore, simulations in a different GCM, GENESIS
\citep{Edsonetal2011,Edsonetal2012}, also show a high albedo for
tidally locked planets. This robustness is due to the simplicity of
the mechanism: any GCM that produces strong convection at the
substellar point should produce a high cloud albedo.

The stabilizing cloud albedo feedback is somewhat sensitive to three
parameters.  First, cloud albedo decreases if liquid cloud
droplets are very large. In CAM3 and CAM4 the cloud droplet size is
a specified parameter based on observations of modern
Earth. % over pristine oceans
If we double its value, the planetary albedo is reduced by
$\approx$0.1.  It is unlikely that the droplet size would be this
high, but it might be possible on a planet with extremely low levels
of cloud condensation nuclei.

%We note that CAM5, which interactively predicts the cloud liquid droplet size,
%produces a cloud liquid droplet size significantly smaller than that
%specified in CAM3 and CAM4 (Table~1).
% It is true, but we do not know whether CAM5 can simulate this well. (JY)

Second, if the orbital period is very short or the planetary radius is
very large, then the Rossby deformation radius is smaller than the
planetary radius and the atmospheric circulation is in the
rapidly-rotating regime
\citep{Edsonetal2011,Showmanetal2013,Leconteetal2013}, which results
in concentrated convection and fewer dayside clouds. Quantitatively,
we find that this regime transition occurs at a 10-day orbit for a
$2R_{\oplus}$ super-Earth, or a 5-day orbit for an Earth-sized
planet. We therefore expect that the substellar cloud feedback we
describe will be weaker for planets orbiting late-M dwarfs.

%The orbital periods at which 
%this occurs are likely too short to correspond to planets near the inner edge 
%of the HZ of red dwarfs, but may be relevant for habitable planets 
%orbiting white dwarfs. 

%Potentially habitable planets around M-stars 
%and K-stars have orbital periods of $\approx$10$-$100 and 
%40$-$120 E-days, respectively, as obtained following Kepler's 3rd 
%Law with the HZ limits of \citet{Kastingetal1993}.
% It seems that the transition occurs at around 10 E-days, right at the inner edge of the HZ. (JY)
% Also our results would decreases the orbital period near the inner edge. (JY)

Third, the planetary albedo can be reduced if a large ocean heat
transport (OHT) from the dayside to the nightside is added to CAM3.
As the OHT is increased, the day--night surface temperature contrast
decreases, weakening atmospheric circulation and producing less
clouds.
%, consequently decreasing the planetary albedo. 
%Increasing OHT also reduces the strength 
%of the nightside temperature inversion and therefore increases 
%the greenhouse effect and surface temperature. 
%Increases in planetary surface temperatures are
%stronger than one might expect based on changes in planetary albedo alone 
If the OHT is similar to that %observed
out of the tropics on modern Earth
\citep[$\approx$10\,W\,m$^{-2}$;][]{TrenberthandCaron2001}, the change
in planetary albedo is negligible. To achieve a reduction in planetary
albedo of $\approx$0.1 requires specifying an OHT ten times that on
modern Earth. This implies that OHT would only disrupt the mechanism
we describe on planets with thick oceans and few continents to disrupt
ocean flows. We confirm this idea using our ocean-atmosphere
simulations with CCSM3.  For a thick ocean without continents, ocean
circulation effectively transports heat from the dayside to the
nightside and the planetary albedo is significantly reduced. The
addition of %even one
obstructing continent(s), however, prevents OHT from disrupting the
cloud feedback so that the planetary albedo is just as high as in CAM3
coupled with an immobile ocean.

%similar to that caused by doubling the cloud droplet size, 
% I comment out it because I want to make the paper as short as possible. (JY)

As addressed in Section~\ref{phases}, substellar clouds could 
be inferred from thermal phase curves because of the suppressed dayside 
OLR. A potential false positive for this effect 
would be a planet with a dry atmosphere %(no moisture or clouds) 
and a high surface albedo region near the substellar point. Such a configuration would reduce the surface temperature and, consequently, the OLR near the substellar point. 
We find that in order to obtain similar dayside OLR 
patterns as those produced by substellar clouds, the substellar surface
albedo must be $\approx$0.8, which is implausible 
for planets around M-stars \citep{JoshiandHaberle2012,Shieldsetal2013}.
 %due to the red stellar spectrum. 
More importantly, such a planet would have much lower
nightside OLR than the moist and cloudy planets considered here. 
% In order to make the paper meets the length limit, I comment out
% this. (JY)
It therefore seems that only clouds can create a significant OLR
minimum at the substellar point in tandem with high nightside
emission.

%%%%%%%%%%%%%%%%%%%%%%%%%%%%%%%%%%%
%%%%%%%%%%%%%%%%%%%%%%%%%%%%%%%%%%%

\section{Conclusions}
We have performed the first 3D global calculations of the effect of
water clouds on the inner edge of the HZ and predict that tidally
locked Earth-like planets have clement surface conditions at twice the
stellar flux calculated by 1D models. This brings already detected
planets, such as HD\,85512\,b and GJ\,163\,c, into the HZ, and dramatically increases
estimates of the frequency of habitable planets. Adopting the planetary
demographics from Figure~19 of \cite{DressingandCharbonneau2013}, our
revised inner edge of the HZ increases the frequency of habitable
Earth-size planets by at least 50--100\%. Crucially, we have also
shown how this stabilizing cloud feedback can be tested in the near
future with thermal phase curves from JWST. 
%If a terrestrial planet with atmospheric water vapor does not 
%exhibit this cloud thermostat, it will suggest that the planet is not tidally locked.

%%%%%%%%%%%%%%%%%%%%%%%%%%%%%%%%%%%
%%%%%%%%%%%%%%%%%%%%%%%%%%%%%%%%%%%

\acknowledgments
We are grateful to D.~Koll, Y.~Wang, F.~Ding, Y.~Liu, C.~Bitz, and
R.~Pierrehumbert for technical assistance and/or  discussions.
D.S.A. acknowledges support from an Alfred P. Sloan Research
Fellowship.
 
 %%%%%%%%%%%%%%%%%%%%%%%%%%%%%%%%%%%
%%%%%%%%%%%%%%%%%%%%%%%%%%%%%%%%%%%

% In order to (1) keep the text as simple as possible and 
% due to (2) the main text has only addressed JWST, so I delete the 
% analyses about MIRI. (JY)

%\footnote{The MIRI spectral range is 5--28~$\mu$m, but the 
%planet/star contrast ratio becomes considerably worse shortward 
%of the planet's blackbody peak. We therefore consider the 
%conservative case of ignoring photons in the 5--10~$\mu$m range.}


%%%%%%%%%%%%%%%%%%%%%%%%%%%%%%%%%
%%%%%%%%%%%%%%%%%%%%%%%%%%%%%%%%%
%%%%%%%%%%%%%%%%%%%%%%%%%%%%%%%%%

%\clearpage

\begin{deluxetable}{lrrrrr}
\vspace{-15mm}
\tablewidth{0pt} %{p{0.8in}  p{0.8in}  p{0.7in} p{0.7in} p{0.7in} p{0.7in} 
%\tablewidth{p{0.8in}  p{0.8in}  p{0.7in} p{0.7in} p{0.7in} p{0.7in} 
\tablecaption{Climate characteristics of terrestrial planets around an M-star}
\tablehead{
\colhead{Group}
&\colhead{Model}  
& \colhead{Experimental Design} 
& \colhead{T$_S$\tablenotemark{a}}
& \colhead{Albedo\tablenotemark{b}}  
& \colhead{$G$\tablenotemark{c}}\\ 
\colhead{}
&\colhead{}  
& \colhead{} 
& \colhead{(K)}
& \colhead{(0--1)}  
& \colhead{(K)}}
\startdata
1\tablenotemark{d}     & CAM3    &   an immobile ocean, S$_0$\,=\,1600\,W\,m$^{-2}$        &  246.7          & 0.48     & 4.9 \\
1     & CAM4    &   an immobile ocean, S$_0$\,=\,1600\,W\,m$^{-2}$        &  246.6         & 0.49     & 6.9 \\
1     & CAM5    &   an immobile ocean, S$_0$\,=\,1600\,W\,m$^{-2}$        &  240.0         & 0.49     & 3.1\\
%\tableline
2\tablenotemark{e} & CCSM3  & aqua-planet, dynamical ocean, S$_0$\,=\,1600\,W\,m$^{-2}$    &  296.2         &  0.26 & 27.6\\
2 & CCSM3  & 1 ridge, dynamical ocean, S$_0$\,=\,1600\,W\,m$^{-2}$           &  264.2         &  0.48 & 20.6 \\
2 & CCSM3  & 2 ridges, dynamical ocean, S$_0$\,=\,1600\,W\,m$^{-2}$         &  263.0         &  0.49  & 20.6 \\
%\tableline
3\tablenotemark{f}    & CAM3     & spin:orbit\,=\,1:1, S$_0$\,=\,1000\,W\,m$^{-2}$  &  222.6  & 0.42 &  7.3 \\
3     & CAM3       & spin:orbit\,=\,1:1, S$_0$\,=\,1200\,W\,m$^{-2}$   &  230.0   & 0.45 & 6.3 \\
3     & CAM3        & spin:orbit\,=\,1:1, S$_0$\,=\,1400\,W\,m$^{-2}$   & 237.8 & 0.47 & 5.6  \\
3     & CAM3        & spin:orbit\,=\,1:1, S$_0$\,=\,1600\,W\,m$^{-2}$   & 246.1    & 0.49 &  6.1 \\
3    & CAM3         & spin:orbit\,=\,1:1, S$_0$\,=\,1800\,W\,m$^{-2}$   & 255.3   & 0.50 & 8.1 \\
3    & CAM3        & spin:orbit\,=\,1:1, S$_0$\,=\,2000\,W\,m$^{-2}$   & 266.2 & 0.51 &  12.2 \\
3   & CAM3          & spin:orbit\,=\,1:1, S$_0$\,=\,2200\,W\,m$^{-2}$   & 280.3  & 0.53 &  21.4 \\
%\tableline
4\tablenotemark{g} & CAM3      & spin:orbit\,=\,2:1, S$_0$\,=\,1000\,W\,m$^{-2}$   & 241.8   & 0.35 &  14.3 \\
4     & CAM3       & spin:orbit\,=\,2:1, S$_0$\,=\,1200\,W\,m$^{-2}$   & 260.3   & 0.34 & 20.1  \\
4     & CAM3       & spin:orbit\,=\,2:1, S$_0$\,=\,1400\,W\,m$^{-2}$   & 270.0  & 0.35 & 20.9 \\
4     & CAM3       & spin:orbit\,=\,2:1, S$_0$\,=\,1600\,W\,m$^{-2}$   & 287.4  &  0.31 & 25.8  \\
%\tableline
5\tablenotemark{h}      & CAM3      & spin:orbit\,=\,6:1, S$_0$\,=\,1000\,W\,m$^{-2}$   &  247.1  & 0.30 &  16.9 \\
5      & CAM3      & spin:orbit\,=\,6:1, S$_0$\,=\,1200\,W\,m$^{-2}$   & 265.4   & 0.28 &  21.9  \\
5      & CAM3      & spin:orbit\,=\,6:1, S$_0$\,=\,1400\,W\,m$^{-2}$   & 284.0   & 0.25  & 27.7 \\
5      & CAM3      & spin:orbit\,=\,6:1, S$_0$\,=\,1600\,W\,m$^{-2}$  &  303.1  & 0.21 & 32.1 \\
%\tableline
6\tablenotemark{i}      & CAM3  &  control (S$_0$\,=\,1200\,W\,m$^{-2}$)  & 230.0 & 0.45  & 6.3  \\
6      & CAM3  &  ice cloud particle size\,$\times$\,0.5                                &  229.1 & 0.47  & 7.0  \\
6      & CAM3  &  ice cloud particle size\,$\times$\,2.0                                & 230.3 & 0.44  &  5.1 \\
6      & CAM3  &  liquid cloud particle size\,=\,3.5\,$\micron$                        & 219.6 & 0.54  &  7.0 \\
6      & CAM3  &  liquid cloud particle size\,=\,7\,$\micron$                        & 223.7 & 0.51  & 6.7  \\
6      & CAM3  &  liquid cloud particle size\,=\,28\,$\micron$                      & 241.8 & 0.34  & 5.1 \\
6      & CAM3  &  cloud parameter $RH_{min}^{hgh}$\,--\,10\% & 229.8 & 0.46 & 6.3  \\
6      & CAM3  &  cloud parameter $RH_{min}^{hgh}$\,+\,10\% & 230.2 & 0.45  & 6.0  \\
6      & CAM3  &  cloud parameter $RH_{min}^{low}$\,--\,10\% & 228.6 & 0.46  & 6.1  \\
6      & CAM3  &  cloud parameter $RH_{min}^{low}$\,+\,10\% & 232.2 & 0.44  & 6.3  \\
6      & CAM3  &  radius\,=\,0.5\,$\times$\,Earth's                                 & 234.7 & 0.40 & 3.1 \\
6      & CAM3  &  radius\,=\,1.0\,$\times$\,Earth's                                 & 230.0 & 0.44 & 3.7  \\
6      & CAM3  &  radius\,=\,3.0\,$\times$\,Earth's                                 & 230.9 & 0.45 & 7.1  \\
6      & CAM3  &  gravity\,=\,4.9 m\,s$^{-2}$                                   & 235.9 & 0.44 & 7.3  \\
6      & CAM3  &  gravity\,=\,9.8 m\,s$^{-2}$                                   & 231.7 & 0.45 & 6.9  \\
6      & CAM3  &  gravity\,=\,19.6 m\,s$^{-2}$                                 & 230.6 & 0.44  & 6.5  \\
6      & CAM3  &  orbital period\,=\,5\,E-days                                   & 248.2 & 0.34  & 14.4 \\
6      & CAM3  &  orbital period\,=\,10\,E-days                                 & 246.1 & 0.36  & 12.0 \\
6      & CAM3  &  orbital period\,=\,15\,E-days                                 & 234.6 & 0.42  & 8.2 \\
6      & CAM3  &  orbital period\,=\,20\,E-days                                 & 230.5 & 0.45  & 8.0 \\
6      & CAM3  &  orbital period\,=\,30\,E-days                                 & 230.6 & 0.45 & 7.6 \\
6      & CAM3  &  orbital period\,=\,100\,E-days                               & 229.6 & 0.46  & 6.1 \\
6      & CAM3  &  mixed layer depth\,=\,1\,m                                                 & 229.0 & 0.45  & 6.1 \\
6      & CAM3  &  mixed layer depth\,=\,10\,m                                               & 229.8 & 0.46  & 6.3 \\
6      & CAM3  &  mixed layer depth\,=\,100\,m                                             & 230.0 & 0.45 & 6.3 \\
6      & CAM3  &  surface pressure\,=\,2\,bars                                                 & 231.4 & 0.45 & 6.7\\
6      & CAM3  &  surface pressure\,=\,5\,bars                                                 & 234.5 & 0.45 & 9.5  \\
6      & CAM3  &  surface pressure\,=\,10\,bars                                              & 237.3 &  0.43  & 8.5 \\
6      & CAM3  &  $P_{(CO_2)}$\,=\,355\,ppmv                                           & 237.5 &  0.45 & 11.0  \\
6      & CAM3  &  $P_{(CO_2)}$\,=\,0.01\,bar                                              & 243.7 &  0.44  & 15.6  \\
6      & CAM3  &  $P_{(CO_2)}$\,=\,0.1\,bar                                                & 252.3 &  0.44  & 21.8 \\
6      & CAM3  &  modern Earth continents I$^{j}$                           & 230.7 &  0.45 & 8.0  \\
6      & CAM3  &  modern Earth continents II$^{k}$                        & 228.5 &   0.46  & 6.6  \\
6      & CAM3  &  dayside--to--nightside OHT\,=\,27\,W\,m$^{-2}$      & 235.4 & 0.44  & 7.9 \\
6      & CAM3  &  dayside--to--nightside OHT\,=\,54\,W\,m$^{-2}$      & 239.5 & 0.41  & 7.5 \\
6      & CAM3  &  dayside--to--nightside OHT\,=\,71\,W\,m$^{-2}$      & 249.5 & 0.38  & 11.5 \\
6      & CAM3  &  dayside--to--nightside OHT\,=\,108\,W\,m$^{-2}$    & 258.8 & 0.36  & 17.0 
\enddata
\tablenotetext{a}{T$_S$ is the global-mean surface temperature.}
\tablenotetext{b}{Albedo is the planetary albedo, primary contributed from clouds.}
\tablenotetext{c}{$G$ is the global-mean greenhouse effect.}
\tablenotetext{d}{Group~1, tidally locked cases, simulated by CAM3, CAM4 and CAM5. For these simulations, the sea-ice modules are switched off because there are significant differences in sea-ice simulation among these models. CAM3 has a resolution of 3.75$^{\circ}$$\times$3.75$^{\circ}$ and 26 vertical levels from the surface to $\approx$30 km. CAM4 and CAM5 have a horizontal resolution of 
1.9$^{\circ}$$\times$2.5$^{\circ}$ and 26 and 30 vertical levels, respectively.}
\tablenotetext{e}{Group~2, tidally locked cases, simulated by CCSM3 without or with meridional barriers. The atmosphere component of CCSM3 is the same as CAM3. The ocean component has a variable latitudinal resolution starting at $\approx$0.9$^{\circ}$ near the equator, a constant longitudinal resolution of 3.6$^{\circ}$, and 25 vertical levels. The CCSM3 simulations have stronger greenhouse effect, this is due to that ocean heat transports from the dayside to the nightside and from the tropics to the extra-tropics of the dayside weaken or eliminate the temperature inversion.}
\tablenotetext{f}{Group~3, tidally locked cases with different stellar fluxes in CAM3. The default 
time step is 1800~s; for high stellar flux, it is reduced to~100~or~50 s to avoid numerical instability.}
\tablenotetext{g}{Group~4, non-tidally locked cases (2:1 spin-orbit resonance) in CAM3}
\tablenotetext{h}{Group~5, non-tidally locked cases (6:1 spin-orbit resonance) in CAM3}
\tablenotetext{i}{Group~6, sensitivity tests for the tidally locked case in CAM3}
\tablenotetext{j}{The substellar point is set to 180$^{\circ}$E over the Pacific Ocean.}
\tablenotetext{k}{The substellar point is set to 20$^{\circ}$E over Africa.}
\end{deluxetable}

\end{document}